\documentclass[sigconf]{acmart}

\usepackage{algorithmic, textcomp, enumitem, multirow, makecell}

\copyrightyear{2025}
\acmYear{2025}
\setcopyright{cc}
\setcctype{by}
\acmConference[SC Workshops '25]{Workshops of the International Conference for High Performance Computing, Networking, Storage, and Analysis}{November 16--21, 2025}{St Louis, MO, USA}
\acmBooktitle{Workshops of the International Conference for High Performance
Computing, Networking, Storage and Analysis (SC Workshops '25), November
16--21, 2025, St Louis, MO, USA}
\acmDOI{10.1145/3731599.3767376}
\acmISBN{979-8-4007-1871-7/2025/11}

\newcommand{\TL}[3][r]{%
  \begin{tabular}{@{}#1@{}}
    #2 \\[-0.8ex] #3
  \end{tabular}%
}

\newcommand{\FloatBodyStyle}{\centering\footnotesize
\renewcommand{\arraystretch}{1.2}
\fontfamily{phv}\selectfont}

\newcommand{\thiswork}{FZModules}

\title[\thiswork]{\thiswork: A Heterogeneous Computing Framework for Customizable Scientific Data Compression Pipelines}

\author{Skyler Ruiter}
\affiliation{
  \institution{Indiana University}
  \city{Bloomington}\state{IN}
  \country{USA}
}
\email{sruiter@iu.edu}

\author{Jiannan Tian}
\affiliation{
  \institution{Oakland University}
  \city{Rochester}\state{MI}
  \country{USA}
}
\email{jtian@oakland.edu}

\author{Fengguang Song}
\authornote{Corresponding author: Fengguang Song, Department of Intelligent Systems Engineering, Luddy School of Informatics, Computing, and Engineering, Indiana University, Bloomington, IN 47408.}
\affiliation{%
  \institution{Indiana University}
  \city{Bloomington}\state{IN}
  \country{USA}
}
\email{fgsong@iu.edu}



\begin{document}

\begin{abstract}
Modern scientific simulations and instruments generate data volumes that overwhelm memory and storage, throttling scalability. Lossy compression mitigates this by trading controlled error for reduced footprint and throughput gains, yet optimal pipelines are highly data and objective specific, demanding compression expertise. GPU compressors supply raw throughput but often hard‑code fused kernels that hinder rapid experimentation, and underperform in rate–distortion. We present FZModules, a heterogeneous framework for assembling error‑bounded custom compression pipelines from high‑performance modules through a concise extensible interface. We further utilize an asynchronous task-backed execution library that infers data dependencies, manages memory movement, and exposes branch and stage level concurrency for powerful asynchronous compression pipelines. Evaluating three pipelines built with FZModules on four representative scientific datasets, we show they can compare end‑to‑end speedup of fused‑kernel GPU compressors while achieving similar rate–distortion to higher fidelity CPU or hybrid compressors, enabling rapid, domain-tailored design.
\end{abstract}

\maketitle

\section{Introduction}\label{fzmod::intro}

Nowadays, large-scale scientific applications and advanced experimental instruments produce extreme-scale data for post hoc analysis and poses unprecedented complexity in data-processing workflows. This leads to not only a high pressure onto supercomputing subsystems (storage, memory, I/O, and communication), but also performance challenges when real-time and inlining processing are attempted. For example, Hardware/Hybrid Accelerated Cosmology Code (HACC)~\cite{hacc} can produce data at an order of magnitude of petabytes in hundred-snapshot simulation of one trillion particles. At the same time, the ``sink'' to receive data, at the moment of data generation, faces overloads due to the excessive data stream-in. To this end, data reduction is becoming an effective method to address the big data issues.

Previous work focuses on building and optimizing data reduction via error-controllable lossy compression as it ensures acceptable reconstruction quality while drastically reducing data size, making it commonly considered for scientific data reduction~\cite{cusz, sz3, sdrbench}.
In many scientific applications, however, CPU based compressors can't reach sufficiently high throughput to be used in these data processing workflows, such as SZ3~\cite{sz3} which as shown in prior work~\cite{pfpl}, achieves tens of gigabytes per second with OpenMP acceleration, leading to the development of GPU accelerated compressors~\cite{cusz, fzg, huang2023cuszp, huang2024cuszp2, zfp}, which often achieve performance an order of magnitude higher than CPU compressors.

On the other hand, when fixing the global tolerable error (``error bound''), while GPU based compressors achieve high throughput, they usually achieve suboptimal rate-distortion performance due to the lower prediction accuracy of high-throughput predictors~\cite{predictor_study}. By contrast, CPU's high-quality predictors involves cascadingly dependent profiling and sampling, hindering the data parallelism and the subsequent high throughput. 
This gap between the throughput and prediction accuracy had led to considerable work to close, such as \textsc{cuSZ}-$i$~\cite{cuszi}, which adapts a lightweight sampling-and-profiling method to optimize the prediction (inspired by QoZ~\cite{HPEZ}), and PFPL~\cite{pfpl}, which implements a portable compressor that can achieve an order of magnitude higher throughput than SZ3 with its OpenMP version. 

With the wide range of data-reduction techniques at use and varying hardware constraints, constructing an optimized compressor to meet the compression ratio, data quality, and throughput for an application requires significant time and effort, as there is no universal best-fit compressor even for a single dataset. Therefore, a modular framework to compose, tailor, and test high-performance compression pipelines is required to facilitate the discovery of powerful custom compressors which meet the performance and quality needs of an application.

In this paper, we present a heterogeneous framework for composing high-performance data reduction modules, FZModules. Built from \textsc{cuSZ}~\cite{cusz}, this expanded framework allows for the rapid development of custom compression pipelines to meet the varied needs of each application and dataset. Specifically, we highlight three pipelines based on previous work: (1) \textbf{FZMod-Default}, which adapts a hybrid CPU-GPU design consisting of a GPU-accelerated Lorenzo predictor followed by CPU-based Huffman encoding, (2) \textbf{FZMod-Speed}, which swaps out the slower Huffman encoder for the efficient bit-shuffle and dictionary encoding of FZ-GPU~\cite{fzg}, and (3) \textbf{FZMod-Quality}, which removes the Lorenzo predictor in favor of the \textit{G-Interp} algorithm found in \textsc{cuSZ}-$i$~\cite{cuszi} for better data prediction. We also support the use of a secondary lossless encoder for further compression, with ZStd~\cite{zstd} being the currently supported secondary encoder module. 
In addition, we experimentally support the rapid creation of compression pipelines utilizing CUDASTF~\cite{cudastf}. This library allows for the description of tasks based on data dependencies, deriving a directed acyclic graph to dynamically manage asynchronous execution of tasks and memory management over the GPU and CPU. 
By utilizing CUDASTF we can leverage task-level concurrency for compression stages not data dependent on each other. FZMod-Default for example, when using its CUDASTF-enabled decompression pipeline, can populate outlier data onto the GPU while the CPU does Huffman decoding on the quantization codes, leveraging the heterogeneous nature of the computing environment.

Our contributions are summarized as follows.

\begin{itemize}[leftmargin=1.3em, topsep=1ex]
    \item We design FZModules, an extensible framework for creating customized error-bounded lossy compressors to meet specific application requirements of speed and reconstruction quality. This framework takes into account the scope of heterogeneous computing systems that make up modern scientific applications to provide high end-to-end speedup while allowing for rapid testing of multiple pipelines. 
    \item We utilize CUDASTF in creating an experimental compression pipeline constructor, which allows for module creation to focus on data needs and high performance. The CUDASTF pipeline automatically leverages derived task concurrency and memory management between tasks, simplifying development and demonstrating the ability of compressors to leverage asynchronous execution on heterogeneous systems. 
    \item We evaluate three FZModules compression pipelines on four real-world application datasets from the Scientific Data Reduction Benchmarks~\cite{sdrbench}. We compare these pipelines against state-of-the-art GPU and CPU based compressors to show our library's ability to attain comparable end-to-end speedup against pure-GPU compressors while also having high rate-distortion performance.
\end{itemize}

FZModules is maintained on GitHub.\footnote{Repository: \url{https://github.com/szcompressor/FZModules}}

We organize the rest of this paper as follows: \S\ref{fzmod::related} discusses related work. In \S\ref{fzmod::design} we present the design, pipeline construction, and CUDASTF functionality. In \S\ref{fzmod::eval} we present the evaluation results for the comparisons between three FZMod pipelines and state-of-the-art CPU and GPU compressors on four real-world datasets. In \S\ref{fzmod::conclusion} we discuss our conclusions and future work.


\section{Related Work}\label{fzmod::related}

\subsection{Scientific Error-Bounded Lossy Compression}\label{fzmod::related::scientific}

Compression of scientific data has long been studied as a way to reduce I/O overhead and optimize data processing workflows. Data reduction techniques fall into either lossless or lossy compression, however, with lossless compression often lacking sufficient compression ratios on scientific floating point data, error-bounded lossy compression has emerged as the best-fit solution. Many techniques have been discovered to facilitate this type of compression, such as the SZ family of compressors~\cite{sz3, qoz} which use a mixture of spline interpolation~\cite{sz3_algo}, linear regression, and Lorenzo extrapolation~\cite{lorenzo}. Transforming is another data-reduction technique commonly used, such as in ZFP~\cite{zfp} and SPERR~\cite{sperr}, followed by their lossless encoding algorithms, or the use of SVD to de-correlate data such as TTHRESH~\cite{tthresh}.

\subsection{GPU-Based Compressors}\label{fzmod::related::gpu}

To meet the ever-increasing performance requirements of scientific applications data generation, GPU compressors have been developed for their massive throughput compared to CPU compressors and the potential to integrate them directly into scientific applications utilizing GPU acceleration. Some examples include \textsc{cuSZ}~\cite{cusz, cuszi, cuszplus}, which FZModules is developed from, that utilizes a Lorenzo predictor or spline interpolator combined with a Huffman encoder. FZ-GPU~\cite{fzg} keeps the Lorenzo predictor developed in \textsc{cuSZ} but replaces the slower Huffman encoder with a bit-shuffle and dictionary encoding technique and a fused kernel to better utilize GPU hardware. cuSZp2~\cite{huang2024cuszp2} optimizes for end-to-end throughput by using a single fused kernel that performs 1D offset prediction and fix-length encoding, creating an exceptionally fast compressor. cuZFP~\cite{zfp} uses a discrete orthogonal transform and attains high ratio and throughput, but doesn't support error-bounded compression only fixed-rate mode. There is also MGARD-GPU~\cite{mgard-gpu} which adapts the CPU based MGARD compressor~\cite{mgard} for GPUs.

\subsection{Composable Pipelines}\label{fzmod::related::composable}

As error-bounded lossy compression has matured, the sheer number of compression algorithms and high-performance data-reduction techniques has led to the development of frameworks to abstract compression stages and pipeline construction. SZ3~\cite{sz3, sz3_algo} creates a modular framework that abstracts the construction of compression pipelines into five stages and allows users to select from provided modules to create custom compression pipelines. SZ3 doesn't include GPU accelerated modules and doesn't provide an extensible framework for creating and testing new modules, resulting in a powerful CPU based modular library with low performance compared to GPU compressors. HPDR~\cite{hpdr} allows for adapting existing compressors using their algorithm abstractions to leverage their framework for better performance scalability, hardware portability, and utilizing overlapping communication and computation to maximize performance in reducing I/O times in supercomputing environments. HPDR focuses on hardware optimizations and powerful tooling to allow for existing compressors to maximize their performance on exascale tasks. Another framework called LC-Framework~\cite{lcframework} also allows for the composition of modules into a pipeline and creation of custom pipelines, as it was used to create PFPL~\cite{pfpl}. LC-framework is powerful, but doesn't capture the heterogeneous nature of modern computing by allowing for mixed CPU-GPU compressor design or leverage asynchronous frameworks. PFPL is a portable error-bounded lossy compressor that has strict enforcement of error bounds and uses an efficient quantizer combined with delta-coding, bit-shuffle, and zero elimination as lossless encoders. 


\section{FZModules: High Level Overview}\label{fzmod::design}

In this section we elaborate on \textsc{cuSZ}~\cite{cusz, cuszi, cuszplus} as the base code which we extended to create FZModules. Next we discuss the modular design of FZModules, how modules can be created and used in pipeline creation. Then we go over the frameworks' composition of modules into a pipeline and explain the role CUDASTF~\cite{cudastf} plays in our experimental pipeline constructor.

\subsection{\textsc{cuSZ} Design}\label{fzmod::design::cusz}

\textsc{cuSZ}~\cite{cusz, cuszplus, cuszi} is a GPU compressor with optional CPU based Huffman encoding for scientific error-bounded lossy compression. Similar to CPU based SZ3, \textsc{cuSZ} employs an abstraction of the compression pipeline into prediction, quantization, and a lossless encoder. In its original adaptation of SZ, \textsc{cuSZ} uses a Lorenzo predictor and then quantizes the prediction error into integer codes or saves data as outliers if they are unpredictable, followed by a GPU Huffman encoder~\cite{cusz}. Further work on \textsc{cuSZ} resulted in the addition of the \textit{G-Interp} interpolation based predictor and an optional secondary lossless codec using NVIDIA Bitcomp~\cite{cuszi}. For more details on the specific data-reduction strategies employed, we refer readers to the \textsc{cuSZ} papers~\cite{cusz, cuszi}.

\subsection{Modularity}\label{fzmod::design::modularity}

Constructing a custom pipeline begins with understanding the available modules and their interactions. Currently, FZModules supports a few modules for each general pipeline stage. For the current preprocessors, the main consideration is whether to use the relative error bounds, resulting in needing to find the data minimum and maximum to normalize the user provided error by the data range. This error type is useful for applying similar error bounds to datasets of varying ranges. The prediction stage supports the multidimensional Lorenzo predictor~\cite{cusz} and the spline interpolator~\cite{cuszi}, with the spline interpolator being a more accurate and slower predictor comparatively. The primary lossless codec can either be an accelerated Huffman encoder or the adapted FZ-GPU~\cite{fzg} bit-shuffle and dictionary encoder. These two encoders have very extreme compression metrics, with the Huffman encoder giving an optimal compression ratio and the FZ-GPU encoder executing significantly faster, but sacrificing compressibility. 
Some modules, such as the Huffman encoder, use global data statistics to support their compression, with the Huffman encoder requiring a histogram of the quantization codes be provided. 
FZModules allows for multiple modules for this type of GPU-accelerated data analysis as well. With the Huffman encoder being able to choose from either a standard GPU-accelerated histogram algorithm or a top-k version, where the top-k outperforms when the distribution of quantization codes has many repeating values. Higher quality-prediction can help generate this data pattern, making the top-k histogram often a better choice for the spline interpolator.
Finally, if the compression ratios are still in need of improvement, a secondary lossless encoder, zstd~\cite{zstd}, can be attempted. 

While there is a low number of total modules currently available, we designed the library to be simple to adapt and update with future modules and provide detailed documentation for those wanting to create and test their own data-reduction modules.

\subsection{Composing Custom Pipelines}\label{fzmod::design::pipelines}

While not as strictly enforced through compile time polymorphism as in SZ3~\cite{sz3}, FZModules continues the common pipeline decomposition into the four main stages: prepossessing, prediction, lossless encoding, and secondary lossless encoding. While not currently containing a high number of individual modules, it is quite simple to construct pipelines with vastly different compression characteristics. As an example, we highlight three compression pipelines:

\begin{itemize}[leftmargin=*]
    \item \textit{FZMod-Default.} This is the default pipeline of FZModules, and consists of a highly parallel Lorenzo predictor and quantizer developed for the original \textsc{cuSZ}~\cite{cusz}, followed by a GPU-accelerated histogram stage, and CPU-based Huffman encoding due to low GPU performance of Huffman encoders. This pipeline attempts to balance throughput, compression ratio, and data quality.  
    \item \textit{FZMod-Speed.} This pipeline attempts to increase the compressor throughput at a cost of lower compression ratio by swapping out the slow Huffman encoder for a faster method of GPU-based bit-shuffle and dictionary encoding adapted from FZ-GPU~\cite{fzg}.
    \item \textit{FZMod-Quality.} To optimize this pipeline for data quality, we swapped out the original Lorenzo predictor for a GPU parallelized interpolation based predictor developed for \textsc{cuSZ}-$i$~\cite{cuszi} which attains higher prediction accuracy, and kept the Huffman encoder for high compression ratio.
\end{itemize}

\subsubsection{CUDASTF Pipelines}\label{fzmod::design::pipelines::cudastf}

As an extension of our custom pipeline creation framework, we added an experimental CUDASTF~\cite{cudastf} pipeline. CUDASTF is a CUDA Sequential Task Flow (CUDASTF) C++ library which allows users to design composable asynchronous algorithms by describing data dependencies between tasks and their execution parameters. This is similar to other libraries which break down algorithms into data spaces and a threading model, such as Kokkos~\cite{kokkos}. CUDASTF uses this information describing tasks in their framework to automatically derive a directed acyclic graph of the dependencies, and executes tasks asynchronously with the needed memory movement between host and device. This powerful tool abstracts away most of the complex programming details, allowing a focus on the core kernel optimizations.

We adapted a subset of the modules to CUDASTF tasks so we can execute a compression pipeline without having to explicitly move data between host and device, and we get task level concurrency for pipelines that can leverage it. Examples of where compressors can leverage task level concurrency is for outlier handling or dynamic module selection based on observed runtime compression results. For outlier handling, a predictor describes data as predictable integer codes or an unpredictable outlier. Then, a task for each type of data can be executed as they no longer share a data dependency until the end of the pipeline when the data is concatenated. In decompression, this can result in one task scattering the outliers to the reconstructed output data from the compressed data, and another task can simultaneously decompress the Huffman encoded data. As this is an experimental design, we avoid performance analysis due to current performance in the tens of gigabytes per second, and instead use it as a way to motivate the power of asynchronous heterogeneous computing in abstract compression pipelines. 


\section{Experimental Evaluation}\label{fzmod::eval}

This section presents our experimental setup and evaluation results. To evaluate FZModules we select three representative pipelines and compare them against state-of-the art GPU and CPU error-bounded lossy compressors on four real-world scientific datasets.

\subsection{Experimental Setup}\label{fzmod::eval::setup}

\begin{itemize}[leftmargin=*]
    \item \textit{Platforms.} We run our experiments on the Indiana University Quartz supercomputer~\cite{quartz}, using its GPU nodes equipped with four NVIDIA Tesla V100 GPUs and its hopper nodes equipped with four NVIDIA H100 GPUs. More details about these platforms are presented in Table \ref{tab:platforms}.
    \item \textit{Baselines.} We compare our FZMod-Default, FZMod-Speed, and FZMod-Quality pipelines against state-of-the-art GPU and CPU compressors, specifically cuSZp2, FZ-GPU, PFPL, and SZ3.
    \item \textit{Test datasets.} Comparisons are done based on four real-world scientific datasets selected from the Scientific Data Reduction Benchmark suite~\cite{sdrbench}. These datasets have been widely used in compression research as they are representative of production level simulations~\cite{cusz, pfpl, hpdr}. More details about the selected datasets can be found in Table \ref{tab:realworld-datasets}.
\end{itemize}

\begin{table}[t]
\caption{Hardware Platforms Used in Experiments}
\label{tab:platforms}
\FloatBodyStyle
\resizebox{\linewidth}{!}{%
\begin{tabular}{@{}lll@{}}
\textbf{Platform} & \textbf{Quartz H100} & \textbf{Quartz V100} \\
\toprule
GPUs & 4-way H100 SXM 80GB & 4-way V100 PCIe 32GB \\
FP32 & 67 TFLOPS & 14 TFLOPS \\
BW & 3.35\,TB/s & 900\,GB/s \\
Driver & 560.35.03 & 560.35.03 \\
CUDA & 12.6 & 12.6 \\
CPUs & 2-way Intel Xeon 6248 & 2-way Intel Xeon 8468 \\
CPU Cores & 40 & 96 \\
RAM & 768\,GB & 512\,GB \\
OS & RHEL 8 & RHEL 8 \\
\TL[l]{Measured}{Bandwidth}~\cite{multi-gpu-bwtest} & $\sim$35.7\,GB/s & $\sim$6.91\,GB/s \\
\bottomrule
\end{tabular}%
}
\vspace{-0.6em}
\end{table}

    \begin{table}[t]
    \caption{Real-world datasets used in the evaluation}
    \label{tab:realworld-datasets}
    \FloatBodyStyle
    \begin{tabular}{@{}lrr@{}}
    \textbf{Datasets} &
    \bfseries\TL{Field Size}{Dimensions} &
    \bfseries\TL{Dataset Size}{\#Fields} \\
    \toprule
    \textbf{CESM-ATM}~\cite{cesm-atm} & \textbf{673.9 MB} & \textbf{20.24 GB} \\[-.6ex]
    \textsc{Climate simulation} & 3600$\times$1800$\times$26 & 33 in total \\
    \textbf{HACC}~\cite{hacc} & \textbf{1.12 GB} & \textbf{6.74 GB} \\[-.6ex]
    \textsc{Cosmology: particle} & 280{,}953{,}867 & 6 in total \\
    \textbf{HURR}~\cite{hurr} & \textbf{100 MB} & \textbf{2 GB} \\[-.6ex]
    \textsc{Hurricane simulation} & 500$\times$500$\times$100 & 20 in total \\
    \textbf{Nyx}~\cite{nyx} & \textbf{536.87 MB} & \textbf{3.22 GB} \\[-.6ex]
    \textsc{Cosmology simulation} & 512$\times$512$\times$512 & 6 in total \\
    \bottomrule
    \end{tabular}
    \vspace{-0.4em}
    \end{table}

\subsection{Evaluation Metrics}\label{fzmod::eval::metrics}

\begin{itemize}[leftmargin=*]
    \item \textit{Compression Ratio.} We compare the compression ratio (CR) of each of the individual compression pipelines and compressors at fixed error bounds of 1e-2, 1e-4, and 1e-6. CR is calculated as the input size divided by the compressed size. All compressors used their value-range-based relative error bound setting, except PFPL which uses more precise definitions for error bound types, meaning relative error bounds for the other compressors is considered Point-Wise Normalized Absolute Error (NOA) for PFPL. This simply means the error bound is calculated as the given error bound divided by the range of the input data.
    \item \textit{Throughput and Overall Speedup.} Compression and decompression throughput of all compressors are measured in GB/s. Overall speedup is a metric proposed in prior work~\cite{aatrox} to consider the improvement a compressor can provide for end-to-end applications, better reflecting real-world use cases. Overall speedup is calculated as:
    \begin{equation}
        \text{speedup} = \frac{1}{\bigl((BW \times CR)^{-1} + T_{\text{compr}}^{-1}\bigr) \times BW} \label{eq:speedup}
    \end{equation}
    where $T_{\text{compr}}$ is compression throughput, $BW$ is memory bandwidth of the medium through which the data is transferred, and $CR$ is compression ratio. This metric accounts for the relationship between CR and throughput, with respect to the speed of the data transfer. For example, when transferring over a 100GB/s network, a compressor with a CR of 2 would need throughput higher than 200GB/s for the compressor to achieve speedup over the given network.
    We use~\cite{multi-gpu-bwtest} to measure GPU-to-CPU transfer times when all four GPUs on the node are reading/writing data and report the results in Table \ref{tab:platforms} as Measured Bandwidth. This allows us to capture the compression performance for a memory system already under load and provides measures closer to observed real-world performance.
    \item \textit{Rate-Distortion} The rate distortion graph shows the compressed bit rate, which is the average bits per input value, and PSNR, to show how well a compressor reconstructs the data with respect to how much it reduces the data size. 
\end{itemize}

\begin{figure*}[tbp] \centering
\begin{minipage}[t][][b]{.8\linewidth}
    \includegraphics[width=\linewidth]{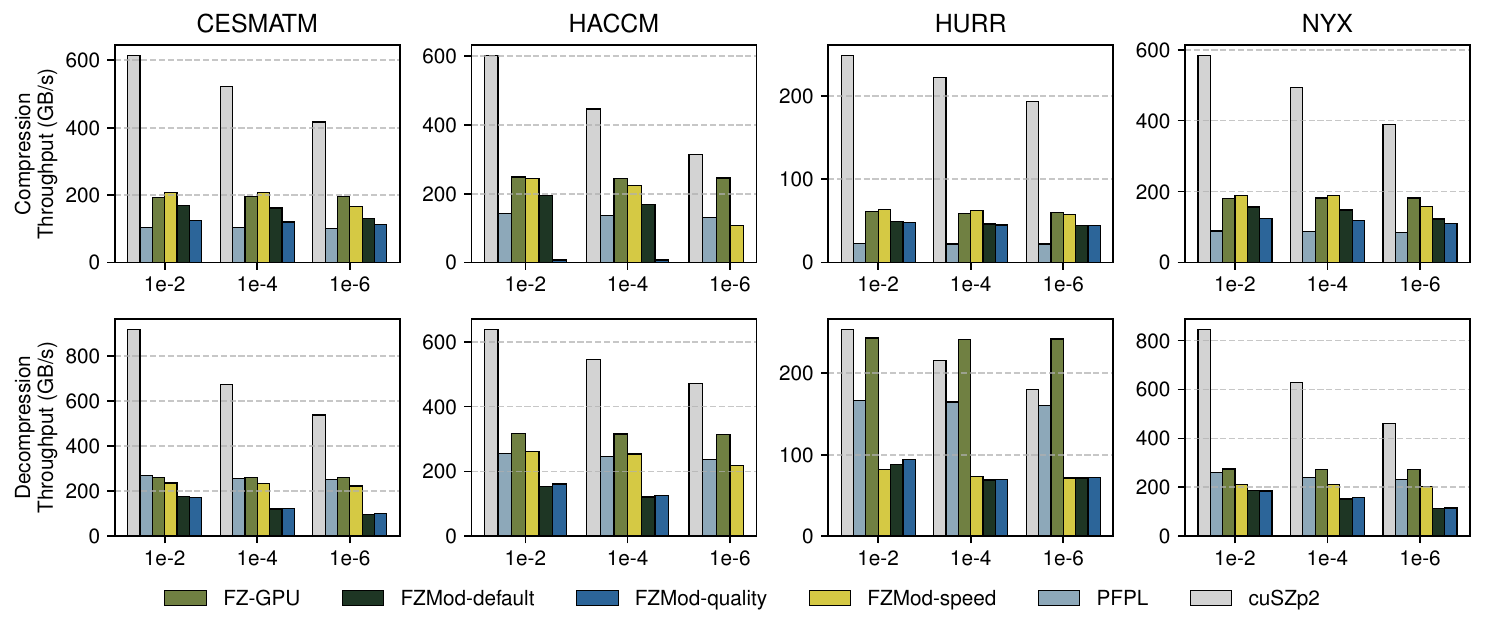}
\end{minipage}\hspace{.01\linewidth}%
\begin{minipage}[t][][b]{.19\linewidth}
    \caption{Compression (top) and decompression (bottom) throughput evaluation on H100 GPU.}
    \Description{Bar plot showing the compression and decompression throughput of the tested compressors on an H100 GPU over the tested error bounds. cuSZp2 excels in throughput with the others falling behind in most cases.}
    \label{fig:throughput_h100}
\end{minipage}
\end{figure*} 

\begin{figure*}[tbp] \centering
\begin{minipage}[t][][b]{.8\linewidth}
    \includegraphics[width=\textwidth]{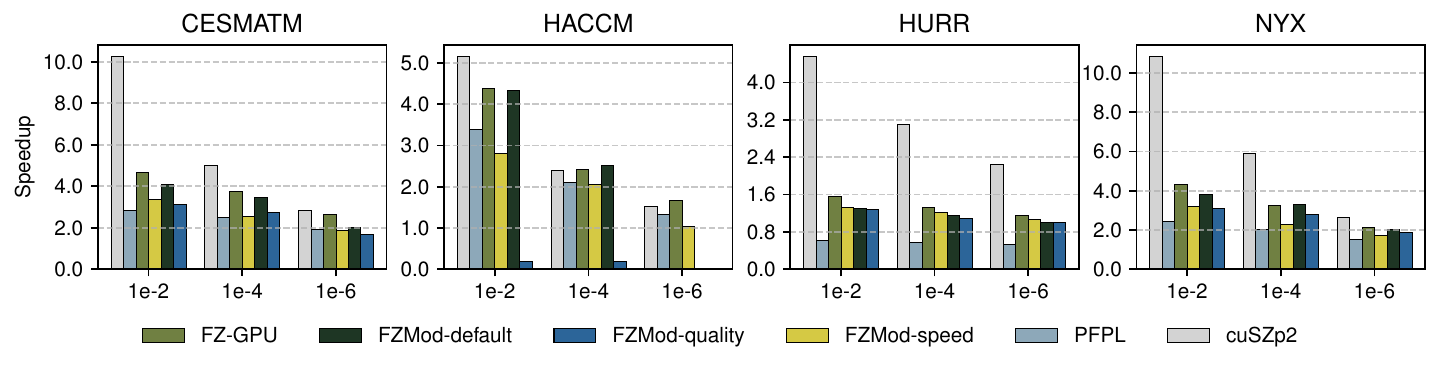}
\end{minipage}\hspace{.01\linewidth}%
\begin{minipage}[t][][b]{.19\linewidth}
    \caption{Speedup evaluation on H100 GPU.}
    \Description{Bar plot showing the overall speedup metric for the tested compressors on the H100 for the tested error bounds. cuSZp2 generally leads in high error bound scenarios and gets brought down to roughly equal with other compressors in tigher error bound situations.}
    \label{fig:speedup_h100}
\end{minipage}
\end{figure*}
\begin{figure*}[tbp] \centering
\begin{minipage}[t][][b]{.8\linewidth}
    \includegraphics[width=\textwidth]{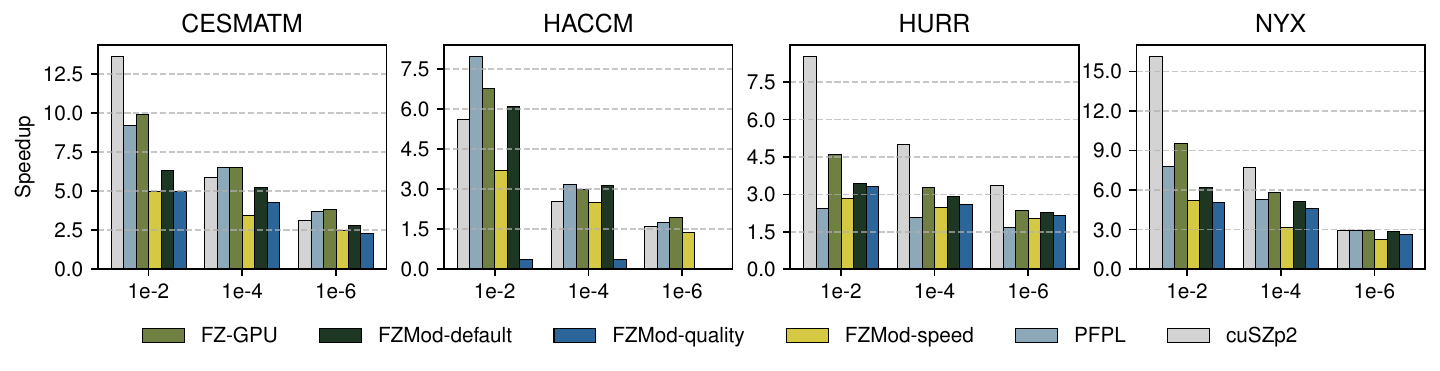}
\end{minipage}\hspace{.01\linewidth}%
\begin{minipage}[t][][b]{.19\linewidth}
    \caption{Speedup evaluation on V100 GPU.}
    \Description{Same plot as above with speedup on the H100, but now with testing done on a V100 GPU. Shows a similar result to H100 plot but with lower GPU throughput to leverage cuSZp2 leads by less of a margin.}
    \label{fig:speedup_v100}
\end{minipage}
\end{figure*}

\subsection{Evaluation Results and Analysis}\label{fzmod::eval::analysis}

\subsubsection{Compression Ratios}

We compressed the datasets using a common range of error bounds and list the results in Table \ref{tab::compression_ratios}. Results for HACC at higher error bounds are excluded from some of the FZModules pipelines due to compression errors in the Huffman encoder. This table shows that SZ3 achieves the highest CR for all datasets and error bounds. This is because of the high quality predictor of SZ3 and shows the high CR that state-of-the-art CPU compressors are able to attain. The second-highest CR is boldfaced in Table \ref{tab::compression_ratios}, where we see PFPL achieves the highest CR of the GPU compressors in 9 of the 12 cases, especially in high error bound scenarios due to its ability to take smooth data and transform it into having long sequences of zeros which are eliminated by its last stage. When the error bound is tightened and the data becomes more difficult to quantize, this allows for the higher accuracy GPU predictors of FZMod-Default and FZMod-Quality to achieve higher or comparable CR to PFPL. 

    \begin{table}[ht]
    \FloatBodyStyle
    \caption{Average Compression Ratios at error bounds 1e-2, 1e-4, and 1e-6. SZ3 has the best compression ratio across the board. Second highest CRs are boldface.}
    \label{tab::compression_ratios}
    \vspace{-1ex}
\resizebox{\linewidth}{!}{%
    \begin{tabular}{@{} >{\bfseries}l l r r r r r r r@{}}
    \bfseries Dataset & \bfseries \textit{eb} &
    \rotatebox{90}{\bfseries \TL[l]{FZMod-}{Default}} &
    \rotatebox{90}{\bfseries \TL[l]{FZMod-}{Quality}} &
    \rotatebox{90}{\bfseries \TL[l]{FZMod-}{Speed}} &
    \rotatebox{90}{\bfseries FZGPU} &
    \rotatebox{90}{\bfseries cuSZp2} & 
    \rotatebox{90}{\bfseries PFPL} & 
    \rotatebox{90}{\bfseries SZ3} \\
    \toprule
    \multirow{3}{*}{\makecell{CESM\\\mbox{-ATM}}} & 1e-2 & 29.9 & 27.7 & 8.4 & 40.5 & 32.6 & \textbf{181.2} & 411.9 \\
                                                  & 1e-4 & 15.8 & 15.0 & 4.9 & 13.0 & 8.3 & \textbf{21.5} & 26.6 \\
                                                  & 1e-6 & 4.8 & 3.9 & 3.2 & 5.4 & 3.8 & \textbf{6.4} & 6.6 \\
    \cmidrule(l){2-9}
    \multirow{3}{*}{HACC}                         & 1e-2 & 22.6 & 5.9 & 5.2 & 12.2 & 7.6 & \textbf{48.7} & 217.9 \\
                                                  & 1e-4 & \textbf{5.6} & 3.2 & 3.1 & 3.7 & 3.0 & 4.9 & 5.8 \\
                                                  & 1e-6 & -- & -- & 1.6 & \textbf{2.2} & 1.8 & 2.1 & 2.5 \\
    \cmidrule(l){2-9}
    \multirow{3}{*}{HURR}                         & 1e-2 & 24.7 & 23.7 & 6.4 & 24.1 & 26.9 & \textbf{76.8} & 475.4 \\
                                                  & 1e-4 & 12.9 & 11.2 & 4.7 & 8.6 & 10.2 & \textbf{17.5} & 34.7 \\
                                                  & 1e-6 & 6.4 & 5.9 & 3.4 & 4.2 & 5.3 & \textbf{8.0} & 13.3 \\
    \cmidrule(l){2-9}
    \multirow{3}{*}{Nyx}                          & 1e-2 & 30.1 & 29.6 & 13.2 & 86.1 & 66.7 & \textbf{1009} & 23038 \\
                                                  & 1e-4 & 18.0 & 20.1 & 4.8 & 16.2 & 22.1 & \textbf{79.4} & 471.5 \\
                                                  & 1e-6 & 6.6 & \textbf{7.4} & 2.8 & 4.0 & 3.7 & 5.6 & 15.9 \\
    \bottomrule
    \end{tabular}%
    }
    \end{table}

\subsubsection{Throughput and Speedup}

The primary metric for many of the pure-GPU compressors is compression and decompression throughput. Due to SZ3 being a low throughput compressor compared to the GPU accelerated compressors, we don't compare it here, and refer readers to ~\cite{pfpl} for a comparison of SZ3 with CPU and GPU compressors. The results of the throughput comparison on an H100 GPU we present in Figure \ref{fig:throughput_h100}. Since cuSZp2 is a throughput optimized fused-kernel compressor, it is natural for it to have the highest throughput among the tested compressors in both compression and decompression throughput, with the gap between cuSZp2 and the second-best compressor being larger on the H100 as takes advantage of the full GPU bandwidth to a higher degree. PFPL and FZ-GPU demonstrates impressive decompression throughput, beating or matching FZModules pipelines in all tested cases. FZMod-Speed uses the same data-reduction techniques as FZ-GPU yet performs worse at times due to not being a fused-kernel implementation. In compression throughput we demonstrate that FZMod-Speed can achieve similar performance to a fused-kernel compressor, FZMod-Quality can outperform PFPL in throughput by anywhere from approximately 20-100\%, and FZMod-Default can fall between the two, with higher throughput than FZMod-Quality and better compression ratio than FZMod-Speed. 

As a way to better capture the performance of these compressors in real-world applications, we evaluate the compressors on overall speedup, taking into account the bandwidth being utilized by the compressed data. We show the results of this comparison for the H100 in Figure \ref{fig:speedup_h100} and for the V100 in Figure \ref{fig:speedup_v100}. These figures demonstrate the extreme throughput of cuSZp2 and the high compression ratios of PFPL bring the compressors much more in line with each other, especially in lower error-bound cases. On the H100 with its significantly higher performance and throughput gives cuSZp2 a clear advantage, yet in the low bandwidth scenario on the V100, PFPL having a much higher CR ends up beating cuSZp2 in overall speedup for 50\% of cases. Of the high data quality compressors, we see that FZMod-Default results in a higher overall speedup than PFPL and FZMod-Quality on the H100 in 8 of 12 cases. This metric helps show how the complex relationship between the compressor algorithm, data characteristics, and hardware result in the best-fit compressor being unique to the application. For example, in scenarios of low-throughput, PFPL could be the best-fit due to its balanced speed and high CR leading to high overall speedup, but if a low error bound is needed and the data is difficult to quantize, a higher accuracy prediction based compressor would likely result in a better overall speedup.

\subsubsection{Data Quality}

    \begin{figure}[!t]
      \centering
      \includegraphics[width=\linewidth]{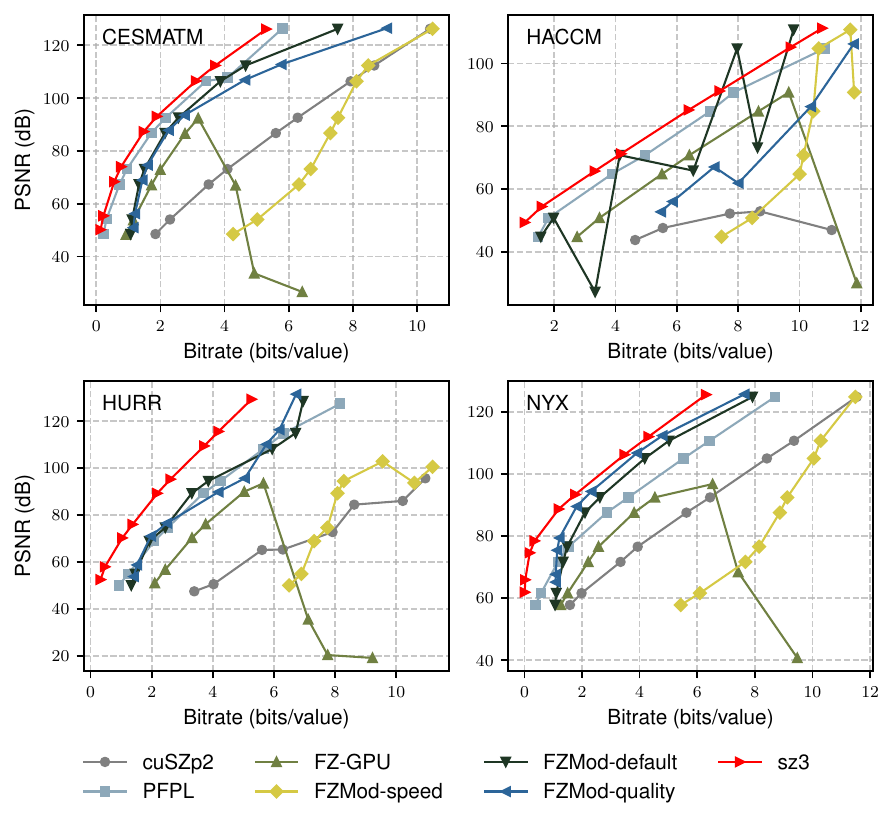}
      \caption{Rate-distortion evaluation.}
      \Description{Line graph showing the rate-distortion curves for each of the tested compressors. Generally shows that the PFPL, SZ3, FZMod-Quality, and FZMod-Default result in all high rate-distortion curves while the others produce poor rate distortion curves.}
      \label{fig:rate_distortion}
      \vspace{-2ex}
    \end{figure}

In cases where the data is generally smooth, and the requirements of the reconstructed data are loose such as in visualization tasks, many general purpose compressors will likely fit the use case as the main consideration then becomes throughput and CR. In many scientific applications, however, the requirements of the application cause many general purpose compressors to fail to meet needed reconstruction quality while maintaining balanced throughput and CR. We see this demonstrated in our evaluation of rate-distortion performance in Figure \ref{fig:rate_distortion}. In this figure we show that the best compressors for rate-distortion are SZ3 as expected, followed by the high-quality compressors of PFPL, FZMod-Default, and FZMod-Quality. The high-throughput compressors of FZ-GPU, cuSZp2, and FZMod-Speed all have poor rate-distortion performance in comparison to the other compressors. This demonstrates the ability of our pipelines to achieve the best, such as on NYX data, or match the best rate-distortion performance of GPU compressors. This demonstrates the ability of our framework to meet many of the use cases requiring high data reconstruction quality while balancing the other compression metrics of CR and throughput, such as for saving simulation snapshots for post-analysis.


\section{Conclusion and Future Work}\label{fzmod::conclusion}


With the massive data generation capabilities of modern scientific applications, error-bounded lossy compression is an essential technique for many data processing workflows. The best-fit compressor for an application, data characteristics, and available hardware is never universal, leading to the vast development of many compressors across GPU and CPU hardware to meet the complex use cases. General purpose compressors of this type suffice for smooth data, high error-bounds, or loose needs on reconstruction quality, but often a custom designed compressor is needed for more complex applications. This paper proposes FZModules, a framework for rapidly building and testing custom compression pipelines to fit the varied needs of scientific applications. By leveraging heterogeneous computing, an experimental sequential task flow pipeline, and a simply designed codebase, we promote the fast construction and testing of arbitrary CPU or GPU high-performance data-reduction modules. By utilizing the adapted modules, we created three compression pipelines, demonstrating compressors with vastly different characteristics that can compare to compressors specialized for a metric. In the future we hope to improve FZModules in the following aspects: (1) Optimize the CUDASTF pipeline to be more performant and have less runtime overhead; (2) Expand the supported GPU and CPU modules and documentation for adding custom modules; (3) Improve the ease of development for custom compression pipelines by developing an auto-selection mechanism for compression modules based on data characteristics, intended hardware environment, and needed quality metrics of the end user.


\begin{acks}

This work was supported in part by the National Science Foundation under Grant Numbers 2311876, 2326495, and 2247060. We also acknowledge the computing resource, Quartz Supercomputer, provided by Indiana University.

\end{acks}

\bibliography{REF}

\end{document}